\documentclass[doublecol]{epl2} 
\usepackage[pdftex]{}

\title{Synchronization of an array of spin torque nano oscillators in periodic applied external magnetic field}
\shorttitle{Synchronization of STNOs in periodic applied magnetic field} %Insert here a short version of the title if it exceeds 70 characters

\author{B. Subash\inst{1} \and V. K. Chandrasekar\inst{1} \and M. Lakshmanan\inst{1}}
\shortauthor{B. Subash \etal}

\institute{                    
  \inst{1}  Centre for Nonlinear Dynamics, Department of Physics, Bharathidasan Univeristy, Tiruchirapalli - 620 024.\\
  
}

\abstract{
Considering an array of spin torque transfer nano oscillators (STNOs), we have investigated the synchronization property of the system under the action of a common periodically driven applied external magnetic field by numerically analyzing the underlying system of Landau-Lifshitz-Gilbert-Slonczewski (LLGS) equations for the macro-spin variables. We find the novel result that the applied external magnetic field can act as a medium to induce synchronization of periodic oscillations, both in-phase and anti-phase, \textit{even without coupling} through spin current, thereby leading to the exciting possibility of enhancement of microwave power in a straightforward way.}

\begin{document}

\maketitle

\section{INTRODUCTION}
It has been increasingly realized that the STNO can be developed into a useful device for submicron  microwave generation due to its varied oscillatory properties as a function of injected spin current and applied magnetic fields\cite{Bertotti:09,kaka,Persson:07}.A rich variety of dynamics has been found in a single STNO by applying a periodically varying spin current\cite{Li:06} or applied magnetic field\cite{Murugesh:09b}.
Even though an individual STNO, which is an interesting nonlinear dynamical system, has practical limitations due to its low output oscillation power\cite{kaka}, nevertheless it has the interesting possibility of generating increased power though the process of synchronization with other similar STNOs under appropriate coupling \cite{Georges:06}. Synchronization of nonlinear oscillators both in the periodic and chaotic regimes are topics of intense discussion in recent times in the nonlinear dynamics literature\cite{7,ML_DV}. For STNOs it has been shown that mutual phase locking phenomenon can be realized leading to increased output power through synchronization in both series and parallel architectures by electrical coupling of spin current with and without delay \cite{Georges:08a,Persson:07}.Although there are various ways of achieving synchronization, even four hundred years ago Huygens observed the same by applying a common load to a system of two pendulum clocks. Recently Nakada, Yakata and Kimura \cite{Nakada:12} have also pointed out the interesting possibility of noise induced synchronization of a pair of STNOs under noisy current injection for controlling the output power in the array of STNOs. This particular study shows that an application of common Gaussian white noise to a system of two uncoupled STNOs acts as a medium for them to induce synchronization.

In this paper, we present an alternative straight forward approach of synchronization of the macro-spin dynamics of an array of STNOs through the simple mechanism of applying a common driven external periodic applied magnetic field in the presence of a dc magnetic field and dc spin current, without any further coupling. The applied periodic magnetic field acts as a medium through which an effective coupling occurs between the macro-spins of the free layers of the different STNOs leading to an effective synchronization involving either in-phase locked or anti-phase locked dynamics. We also present a phase diagram for a system of two STNOs delineating the in-phase and antiphase synchronization and desynchronization regions in the dc magnetic field vs dc spin current plane. We also present the results for 4 and 100 STNOs to confirm the possibility of synchronization of a large number of STNOs, even in the presence of mismatches in system parameters.

\section{Nonlinear dynamics of a Spin Torque Nano Oscillator}
The dynamics of the magnetization in the free layer \cite{Slonczewski:96,Berger:96} of an STNO is known to be well described by the LLGS equation for the normalized spin vector $\vec{S}$, which is a classical unit vector, $\vec{S}=S_{x}\hat{i}+S_{y}\hat{j}+S_{z}\hat{k}$, $|\vec{S}|^{2}=1$. It is given by the nonlinear evolution equation, see for example ref.\cite{Laksh:12},

\begin{equation}
\frac{d\vec{S}}{dt}=-\gamma\vec{S}\times\vec{H}\,_{eff}+\lambda\vec{S}\times\frac{d\vec{S}}{dt}-\gamma a\vec{S}\times(\vec{S}\times\hat{S_{p}}),
\label{eqn1}
\end{equation}
where
\begin{equation}
\vec{H}_{eff}=\vec{H}_{exchange}+\vec{H}_{anisotropy}+
\vec{H}_{demag}+\vec{H}_{applied}.
\label{eqn3}
\end{equation} 
In the above $a$ is the density of the spin current, defined as 
$
a={\hbar \eta j}/{2S_0Ve}
$, $\eta$ is the area of cross section of the pillar, $j$ is the charge current density, $S_0$ is the saturation magnetization, $V$ is the volume of the fixed layer. It is to be noted that the the density of the spin current is proportional to the density of the charge current injected into the fixed layer. Note that with the above definition $a$ can be expressed in the unit of Oersted\cite{Li:06}. Hence by varying the charge current one can vary the spin current injected into the free layer. 
\begin{figure}[!h]
\centering \includegraphics[width=1.0\linewidth]{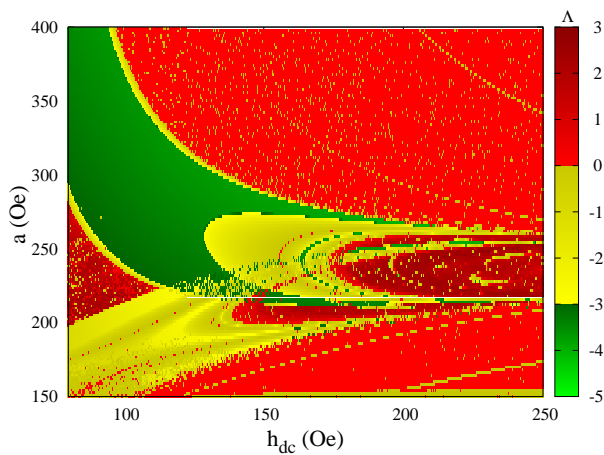} \caption{The largest Lyapunov exponent ($\Lambda$) is plotted in the $a$-$h_{dc}$ plane for a single STNO for an oscillating external magnetic field of strength $h_{ac}=10 $~Oe and frequency $\Omega=15 $~GHz with anisotropy parameter $\kappa=45.0$~Oe along the in-plane axis.  
\revision{   The colour coding clearly distinguishes periodic and chaotic regimes.}  \label{bif} }
\end{figure}

Considering homogeneous spin states on the ferromagnetic free layer, we can treat $\vec{H}_{exchange}$ to be zero. We assume an easy plane anisotropy, $\vec{H}_{anisotropy} = (\kappa S_x,0,0)$ where $\kappa$ is the strength of the anisotropy. For a nano pillar geometry, we assume $\vec{H}_{demag}=-4\pi S_0(0,0,S_z)$ and $\vec{H}_{applied}=(\mathcal{H}(t),0,0) $ so that the external magnetic field is applied along the $x$-direction. In our following analysis, as illustration, we choose the parameters typically as that of a permalloy thin film. So we have fixed the Gilbert damping parameter as $\lambda=0.02$, saturation magnetization as $4\pi S_0 = 8.4~ $ {kOe}, and the gyromagnetic ratio as $\gamma=0.01767~$ Oe$^{-1}$ ns$^{-1}$.

%In our further analysis the Gilbert damping ($\lambda=0.02$), saturation magnetization($4\pi S_0 = 8.4~ \mbox{kOe}$) and the gyromagnetic ratio($\gamma=0.01767~ \mbox{Oe}^{-1} \mbox{ns}^{-1}$) are chosen for the case of permalloy thin film. 
We assume the spin polarization vector $\vec{S}_p = (1,0,0)$ and that the applied external magnetic field consists of dc and ac components. Therefore,

\begin{equation}
\vec{H}_{eff}= \mathcal{H}\hat{i}+\kappa S_x\hat{i} -4\pi S_0 S_{z}\hat{k}, \mathcal{H}(t)=h_{dc}+h_{ac} \cos(\Omega t).
\end{equation}

Under a stereographic projection, eq.~(\ref{eqn1}) can be equivalently rewritten using the complex scalar function \cite{Lakshmanan:84}

\begin{eqnarray}
\omega(t)&=&\frac{S_{x}+iS_{y}}{1+S_{z}}, \\ 
S_{x} = \frac{\omega + \bar{\omega}}{1+|\omega|^2}~;~~
S_{y}& = &-i\frac{(\omega - \bar{\omega})}{1+|\omega|^2}~;~~
S_{z} = \frac{1-|\omega|^2}{1+|\omega|^2}.\nonumber
\label{eqn2}
\end{eqnarray} as

\begin{eqnarray}
(1-i\lambda)\frac{d\omega}{dt}&=&\gamma(a-\frac{i\mathcal{H}}{2})(1-\omega^2) \nonumber \\
 &-& \frac{1}{2}i\gamma\kappa\frac{\omega+ \omega^*}{ 1+\omega\omega^*}(1-\omega^2) \nonumber \\
 &-& i\gamma4\pi S_{0}\frac{1-\omega\omega^*}{1+\omega\omega^*}\omega
\label{w}
\end{eqnarray}

One may note that the effect of applied magnetic field and the spin current terms occur together in the first  term in the right hand side of eq.~(\ref{w}). So one may expect the effect of spin current and magnetic field to complement each other. It has been shown\cite{Yang:07,Murugesh:09b} that in the case of homogeneous magnetization in the free layer, where the spin $ \vec{S}$ is a function of time only, $ \vec{S} = \vec{S}(t)$, LLGS equation can exhibit interesting bifurcation and chaos scenario depending on the  strength and form of the applied external magnetic field and injected current.\revision{ In fig.~\ref{bif}, we present a two parameter ($a-h_{\mbox{dc}}$) phase-like diagram characterized by the value of the largest Lyapunov exponent which especially distinguishes periodic and chaotic regimes. This is }obtained numerically by integrating eq.~(\ref{eqn1}) or eq.~(\ref{w}) of a single STNO using embedded variable step size Runge-Kutta method in the current ($a$) Vs applied magnetic field ($h_{dc}$) plane with anisotropy strength $\kappa=45$~Oe,where the other parameters are chosen as given in the figure caption. Note that the spin current ($a$) and the applied magnetic field ($h_{dc}$) are in units of Oersted(Oe).

 All our numerical analysis have been done on a 32-node cluster facility with 8 digit accuracy in the normalization of the magnetization vector $\vec{S}$. In fig. \ref{bif} the 
\revision{   yellow/green} regions are periodic, where all the Lyapunov exponents are zero or negative. In the 
\revision{   red} region one of the Lyapunov exponents(among the three) is greater than zero, that is  positive as their values are encoded in color palette. One may note that in the lower ranges of magnetic field and current strength limit cycles/periodic oscillations are present, whereas in the higher strength regions the spins are in chaotic motion.
One can also check that typical transition from periodic to chaotic states occur through well defined routes such as period doubling bifurcations\cite{Murugesh:09b}.
% as a coefficient function in front of the quadratic term in $\omega$

\section{Synchronization of coupled STNOs}
\begin{figure}[!h]
 \centering \includegraphics[width=1.0\linewidth]{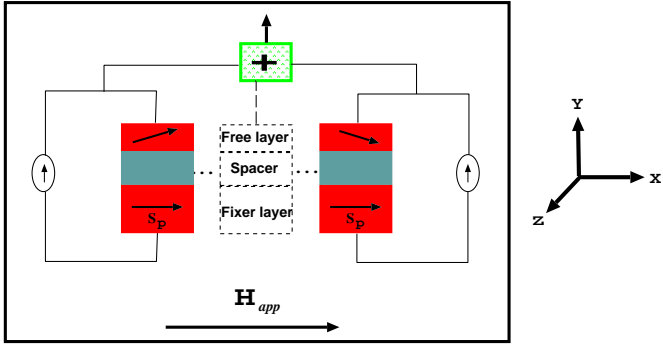} \caption{Schematic representation of an array of two STNOs placed in the oscillatory external magnetic field ($\vec{H}_{app}$). The $+$ sign stands for summing up of two output currents. \label{model} }
\end{figure}

The schematic representation of our model for an array of two STNOs (see fig.~\ref{model}) in the presence of a common applied magnetic field 
\begin{equation}
\vec{H}_{app}=(h_{dc}+h_{ac}\cos \Omega t,0,0),
\label{app}
\end{equation}
where each one of the STNOs is having a separate dc charge current source and the synchronized output current from each STNO as taken in its free layer is added as shown in fig.~\ref{model}. Considering a  system of two electrically uncoupled STNOs placed in a common magnetic field(\ref{app}), the magnetization dynamics is governed by the LLGS equations for the magnetization vectors $\vec{S_{1}},\vec{S_{2}},$ 
% \begin{widetext}

%\begin{equation} 

{\small
\begin{eqnarray}
\frac{d\vec{S_{1}}}{dt} & = & -\gamma\vec{S_1}\times
\vec{H_1}_{\mbox{eff}}+\lambda\vec{S_1}\times\frac{d\vec{S_1}}{dt} 
-\gamma a\vec{S_1}\times(\vec{S_1}\times\hat{S_{p}}), ~~~~\nonumber\\
\frac{d\vec{S_{2}}}{dt} & = &-\gamma\vec{S_2}\times
\vec{H_2}_{\mbox{eff}}+\lambda\vec{S_2}\times\frac{d\vec{S_2}}{dt}-\gamma a\vec{S_2}\times(\vec{S_2}\times\hat{S_{p}}) ~~~~
\label{coupled}
\end{eqnarray}
}

%\end{equation} 

%\end{widetext}
\begin{figure}[!ht]
\begin{center}
\includegraphics[width=1.0\linewidth]{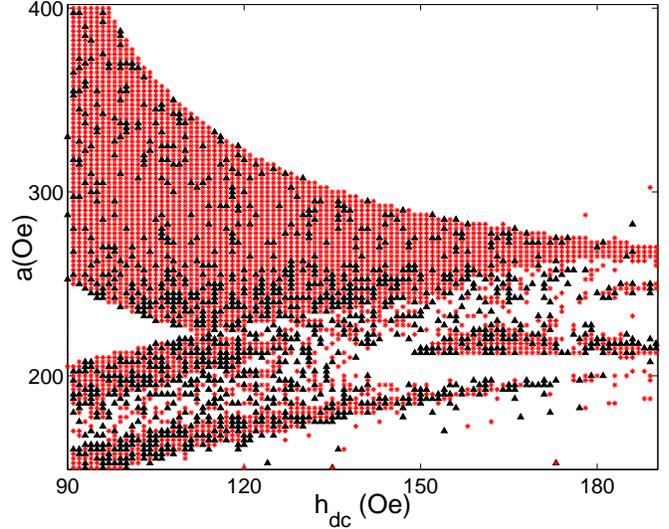}
\caption{The in-phase(red) and anti-phase(black) synchronization regions of two similar STNOs having same anisotropy field $\kappa = 45.0$~Oe placed in an oscillating external magnetic field of strength $h_{ac}=10$~Oe and all the other parameters same as in fig.~\ref{bif}. White regions correspond to  desynchronization \label{bif_io} }
\end{center}
\end{figure}

or equivalently the corresponding evolution equations for the stereographic variables $\omega_1(t)$ and $\omega_2(t)$. Here $\vec{H_1}_{eff}$ and $\vec{H_2}_{eff}$ are of the same form as $\vec{H}_{eff}$ given by eq.~(3), except that the spin variable $\vec{S}$ in (3) is replaced by $\vec{S}_1$ and $\vec{S}_2$ in $\vec{H_1}_{eff}$ and $\vec{H_2}_{eff}$, respectively, with all the other terms remaining the same as in eq.~(3). Thus both the STNOs are under the influence of the common applied field given by eq.~(6). We numerically integrate the above set of equations for a range of parameters in the $h_{\mbox{dc}}$ Vs   $a$ plane \revision{   with random initial conditions} for fixed values of oscillating magnetic field strength $h_{\mbox{ac}}$  and frequency $\Omega$, indicating regions of synchronization of periodic oscillations (limit cycles), both in-phase (red) and anti-phase (black),  as well as desynchronization regions 
\revision{  (which includes both periodic and chaotic regimes) as shown in fig.~\ref{bif_io}.  In fig.~\ref{bif_io} the strong mixing of in-phase and anti-phase synchronizations in the ($a- h_{\mbox{dc}}$) plane is essentially due to the choice of random initial conditions in our numerical analysis.}
\begin{figure}[!ht]
\begin{center}
\includegraphics[width=0.7\linewidth]{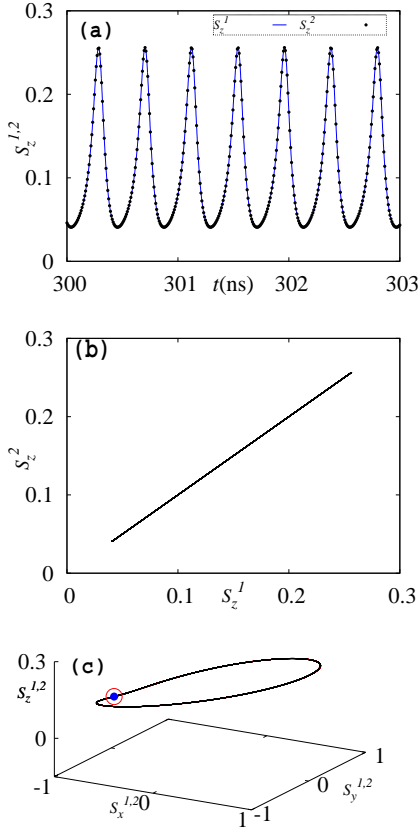}
\caption{Synchronization dynamics for an array of two identical STNOs having same anisotropic field $\kappa = 45.0$~Oe placed in the oscillating external magnetic field of strength $h_{ac}=10$~Oe of frequency $\Omega=15~ns^{-1}$ , $h_{dc}=130 $~Oe and $a=300 $~Oe. (a) time series plots of $S_{z}^{1} $ and $ S_{z}^{2}$ (b)phase space plot of z component($S_{z}^{1,2}$) of the magnetization vectors. (c) the orbits  of the magnetization vectors $\vec{S}^{1}(t)$ and $\vec{S}^{2}(t)$, showing complete in-phase synchronization. Here the open and closed circle points are the values of $\vec {S^{1}}$ and  $\vec {S^{2}}$ at a particular instant of time(t ns). \label{sameip} }
\end{center}
\end{figure}

Specifically in fig.~\ref{sameip} we describe the synchronization dynamics of the spin vectors $\vec{S_{1}}$ and $\vec{S_{2}}$ for two identical STNOs placed in a common external magnetic field. Fig.~\ref{sameip}(a) depicts the time series plots of signals of the $z$-components of both the spin vectors, while fig.~\ref{sameip}(b) presents the phase space structure ($S_{z}^{1} - S_{z}^{2}$) of these components. Fig.~\ref{sameip}(c) displays the orbits of the two spin vectors themselves. We have chosen the system parameters specifically with the choice of anisotropic field strength $\kappa=45$~Oe,  density of the spin current $a=220$~Oe, dc component of the magnetic field, $h_{dc}=130 $~Oe, with an oscillating magnetic field of strength $h_{ac}=10$~Oe and frequency $\Omega=15~ns^{-1}$. The complete coincidence of the time series plots and orbits of the spin vectors along with the $45\,^{\circ}\mathrm{}$ diagonal line in the $S_{z}^{1} - S_{z}^{2}$ phase space clearly confirms the complete in-phase synchronization of the magnetization vectors of the two identical STNOs.
%Specifically in fig.~\ref{same}, we plot the $S_x$, $S_y$ and $S_z$  components of the spin vectors of two identical oscillators eq.~(\ref{coupled}) for the anisotropic field strength $\kappa=45$~Oe, both time series and phase space plots. 
%In fig.~\ref{same}a, the dotted line is the time series of the x component($S_x$) of the first oscillator magnetization vector($\vec{S_{1}}$) and solid line corresponding to that of the second oscillator($\vec{S_{2}}$ ). Similarly plotted for the other two components of the $\vec{S_{1,2}}$. fig.~\ref{same}d, is the phase space plot of the z component($S_z$) of the two oscillators having $a=220$~Oe clearly shows the existence of in-phase synchronization.  In fig.~\ref{same}e, the $S_{x}^2$(z component of $\vec{S_{2}}$) is plotted with dotted line to differentiate  from that of the $S_{x}^1$, along with the phase space diagram fig.~\ref{same}h, it is clear that the system is exhibiting in-phase synchronization for the value of spin current density $a=221$~Oe for the same anisotropy. 

\begin{figure}[!ht]
\begin{center}
\includegraphics[width=0.7\linewidth]{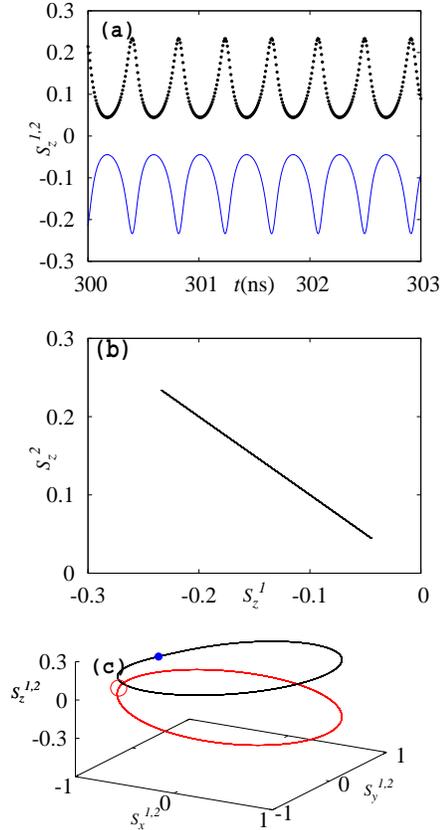}
\caption{Anti-phase synchronization of a system of two STNOs with the choice of parameters $h_{dc}=106 $~Oe and $a=250 $~Oe and all the other parameters same as fig.~\ref{sameip}. (a)time series plot of  $z$-components (${S_z^{1,2}}$), (b) $(S_z^1 - S_z^2)$  phase space structure. (c) the spin orbits, where $\vec {S^{1}}$ (open circle) and  $\vec {S^{2}}$ (closed circle) at a particular instant of time (t=301.5 ns) is indicated. \label{sameap} }
\end{center}
\end{figure}

For a slightly different choice of parameters namely $h_{dc}=106 $~Oe and current density $a=250 $~Oe with all the other parameters remaining the same as in fig.~\ref{sameip}, we obtain an anti-phase synchronization of the two spin vectors, as depicted in fig.~\ref{sameap}, where the $y$ and $z$ components are in anti-phase state while the $x$-component is in an in-phase state (which is allowable due to the constraint, $\vec{S}^{2}=1$). Note that since the direction of the spin current is along the $y$-direction (fig.~\ref{model}) and the polarization of the spins is along the $x$-direction, the anti-phase synchronization along the $y$ and $z$ directions does not alter the resistance to the spin current. So both in-phase and suitable anti-phase synchronizations are equally desirable dynamical states.

\begin{figure}[!ht]
\begin{center}
\includegraphics[width=0.7\linewidth]{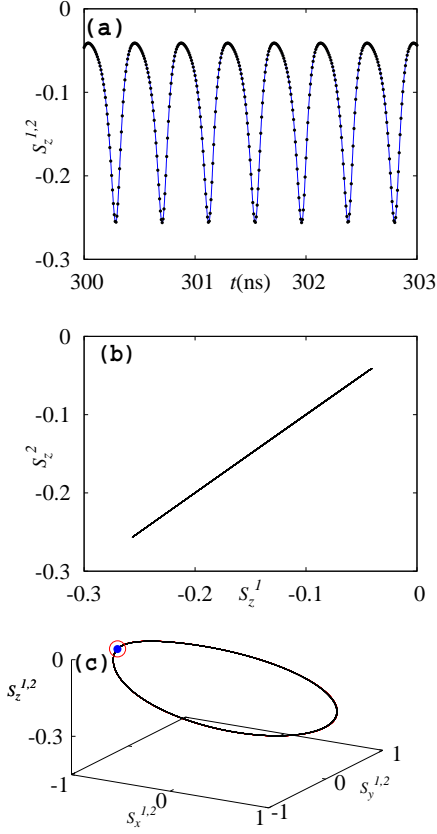}
\caption{Synchronization of an array of two different STNOs with anisotropy fields $\kappa_{1} = 45.0 $ Oe and $\kappa_{2} = 46.0 $~Oe placed in the oscillating external magnetic field, $h_{dc}=130 $~Oe and $a=300 $~Oe with all the other parameters remaining same as fig.~\ref{sameip}. \label{diff} }
\end{center}
\end{figure}

Of course real STNO devices cannot be exactly identical and the synchronization has be to robust against mismatches in system parameters to a reasonable extent. In order to check the robustness of the above synchronization of two STNOs we consider the dynamics of eq.~(\ref{coupled}) with different anisotropic field strengths, $\kappa_1=45.0$~Oe and ~$\kappa_2=46.0$~Oe for the systems 1 and 2 repectively, with other parameters chosen as in fig.~\ref{sameip}. The results are depicted in fig.~\ref{diff}. Again we find that the oscillators are in complete in-phase synchronization. In fact we have checked the existence of robust synchronization up to a choice of $\kappa_2=55.0$~Oe.

\begin{figure}[!ht]
\begin{center}
\includegraphics[width=0.7\linewidth]{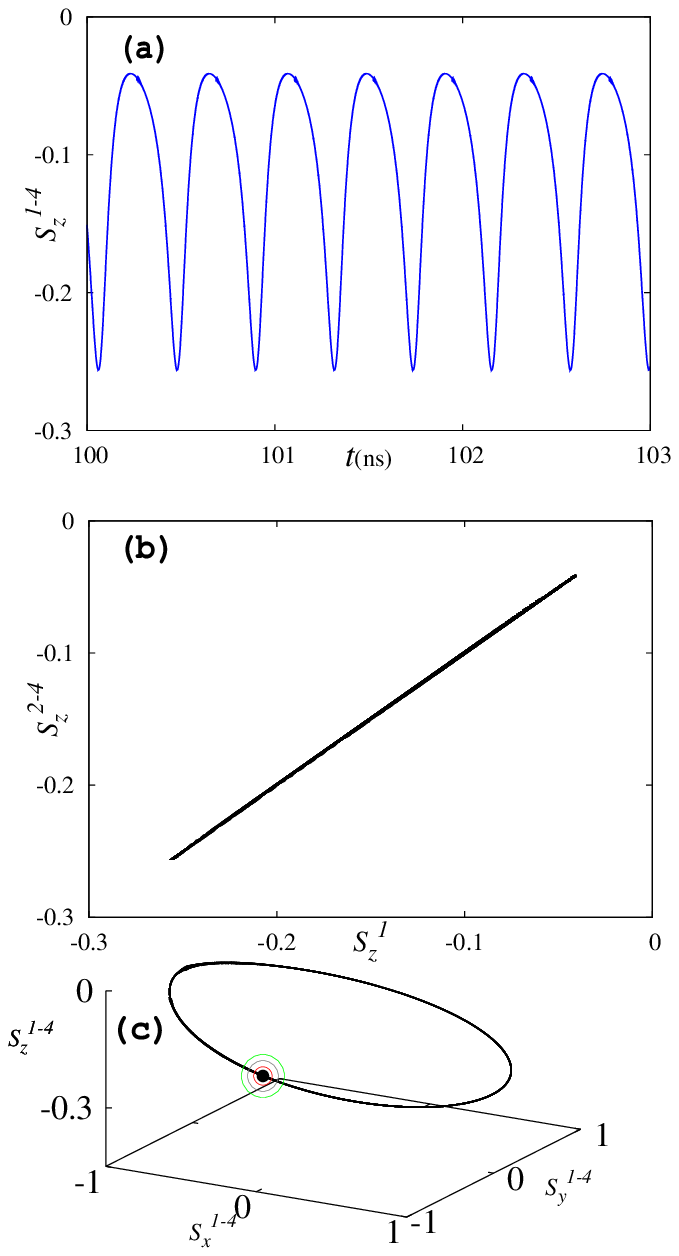}
\caption{Synchronization dynamics of an array of four different STNOs having anisotropy fields $\kappa_1 = 45, \kappa_2=46, \kappa_3=47$ and $\kappa_4=48 $ Oe, $h_{dc}=130 $~Oe and $a=300 $~Oe with other parameters chosen as in fig.~\ref{sameip}.\label{diff4} }
\end{center}
\end{figure}

Next we have extended our analysis to an array with larger number of oscillators. We have verified the above synchronization features for N = 4,6,10 and 100 oscillators. For N=4 in fig.~\ref{diff4} we have depicted the synchronization dynamics for the choice of anisotropy parameters $\kappa_1=45.0, ~ \kappa_2=46.0,~ \kappa_3=47.0 $ and $ \kappa_4=48.0 $~Oe with the remaining parameters chosen as in  fig.~\ref{sameip}. Here fig.~\ref{diff4}a presents the time series of the $z$-components of all the four spins, while in fig.~\ref{diff4}b the diagonal line represents the collective phase space structure  ($S_{z}^{1} - S_{z}^{2-4}$). In fig.~\ref{diff4}c we have drawn the orbits of all the four spin vectors to demonstrate the in-phase synchronization behaviour in the case of four STNOs also.

\begin{figure}[!ht]
\begin{center}
\includegraphics[width=0.7\linewidth]{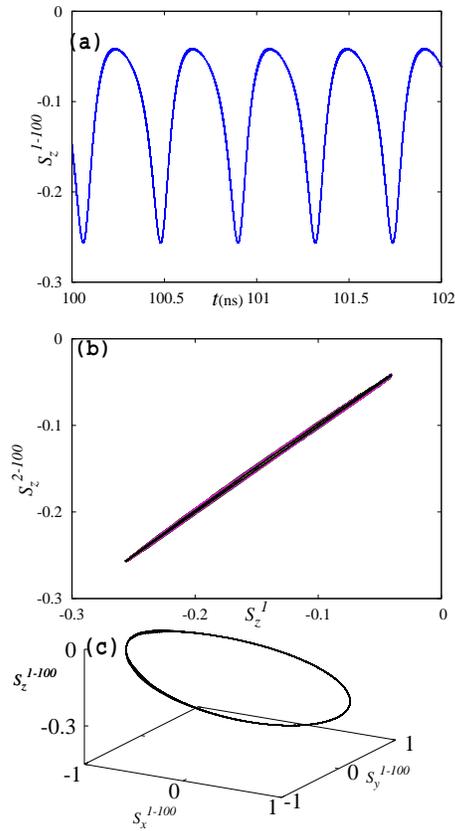}
\caption{The time series (a) and phase space (b) plots of an array of 100  nonidentical STNOs showing the in-phase synchronization for external magnetic field of strength $h_{dc} = 130$~ Oe, external current $a = 300$~ Oe and anisotropy strength $\kappa_i$, i = 1,2,...100 distributed randomly between 45 to 55 Oe. \label{100ip} }
\end{center}
\end{figure}

Finally in fig.~\ref{100ip} we have presented the spin dynamics of a system of 100 STNOs in the presence of a dc spin current and combined dc and ac common external applied magnetic field of strength and frequency as in fig.~\ref{sameip} but with anisotropy parameters $ \kappa_i,i=1,2,....,100$, distributed randomly between $45-55$~Oe. Again one can identify a satisfactory almost complete in-phase synchronization. We have verified the synchronization dynamics even for a random choice of $ \kappa_i$'s between 45 and 80~Oe as well, after which desynchronization sets in a slow manner (that is, a certain number of oscillators are synchronized, while the others are desynchronized). We have also checked the existence of anti-phase synchronization for suitable choices of parameters and a combined in-phase--anti-phase synchronization for yet another set of parameters. The full details will be presented elsewhere.

\section{SUMMARY AND CONCLUSION} 
In this paper we have shown the exciting possibility of synchronization of STNOs in the presence of a periodically varying applied driven magnetic field in addition to constant dc magnetic field and dc spin current but without any coupling, so that the applied magnetic field acts as a medium to induce synchronization of periodic oscillations. This fact coupled with the already proven possibility of synchronization due to coupling through the spin current of STNOs with or without delay or due to the presence of common noise, or mass synchronization of groups of coupled Kuramoto type oscillators\cite{Sheeba:10}, can help to make further progress in the experimental realization towards practical usage of STNOs. Further, we also note that the recent experimental study on the action of applied magnetic field on a single nanowire\cite{magfield:03} also can help the possibility in realization of such a mass synchronization in STNOs.

\acknowledgments
The work forms part of a Department of Science and Technology (DST), Government of India, IRHPA project and is also supported by a DST Ramanna Fellowship of M. L. He has also been financially supported by a DAE Raja Ramanna Fellowship.

%\newpage %Just because of unusual number of tables stacked at end

%\newpage %Just because of unusual number of tables stacked at end

\end{document}